\documentclass[aps,prl,twocolumn,amsmath,amssymb,superscriptaddress]{revtex4}

\usepackage{graphicx}       
\usepackage{dcolumn}        
\usepackage{bm}             
\usepackage{amsmath}        
\usepackage{color}          

\DeclareMathOperator{\sgn}{sgn}

\definecolor{parchment}{rgb}{.9,.9,.6}

\newcommand{\bq}{\bm q}
\newcommand{\bQ}{\bm Q}
\newcommand{\bk}{\bm k}
\newcommand{\br}{\bm r}

\begin{document}
\title{Magnetic field evolution of the quasiparticle interference in a
$d$-wave superconductor}
\author{T. Pereg-Barnea}
\affiliation{Department of Physics, University of Texas at Austin, Austin TX 78712-1081, USA}
\author{M. Franz}
\affiliation{Department of Physics and Astronomy, University of British Columbia, Vancouver, B.C. V6T 1Z1, Canada}
\date{\today}
\begin{abstract}
Quasiparticle interference in a $d$-wave superconductor with weak disorder
produces distinctive peaks in the Fourier-transformed local
density of states measured by scanning tunneling spectroscopy. We predict
that amplitudes of these peaks can be enhanced or suppressed by applied
magnetic field according to
a very specific pattern governed by the symmetry of the
superconducting order parameter. This calculated
pattern agrees with the recent experimental
measurement and suggests that the technique could be useful for probing
the underlying normal state at high fields.
\end{abstract}

\maketitle


There now remains little doubt that hole-doped cuprate superconductors below
their critical temperature $T_c$ form a rather conventional BCS superconducting
(SC) state characterized by a spin-singlet $d$-wave order parameter. The key
mystery in the field is the nature of the state that occurs when
superconductivity is suppressed by underdoping, magnetic field, or
temperature \cite{franz0}. Recently observed quantum oscillation
phenomena in high magnetic fields \cite{taillefer1} indicate a metallic
state with small Fermi pockets which appear incompatible with the standard
band structure calculations. These results are also difficult to reconcile
with the angle-resolved photoemission (ARPES) data \cite{kanigel1,shen1},
which instead imply ``Fermi arcs'', i.e.
disconnected segments of a Fermi surface that appear above $T_c$ close to the
nodal points of the $d$-wave order parameter.

It is possible, in principle, that the normal state reached by the
applied magnetic field is different from the state above $T_c$. However,
since ARPES cannot be performed in high magnetic fields and quantum oscillations
are difficult to detect at elevated temperatures, another technique is
needed to settle this puzzle. A good candidate is the scanning
tunneling probe which can be applied
at finite temperature as well as in the presence of magnetic
fields \cite{fischer1}. The recently perfected technique of
Fourier-transform  scanning tunneling spectroscopy (FT-STS) allows extracting
the Fermi surface from the dispersion of the quasiparticle interference
peaks \cite{Hoffman,mcelroy1,howald1}
and yields results in good agreement with ARPES.

\begin{figure}
\includegraphics[width=8cm]{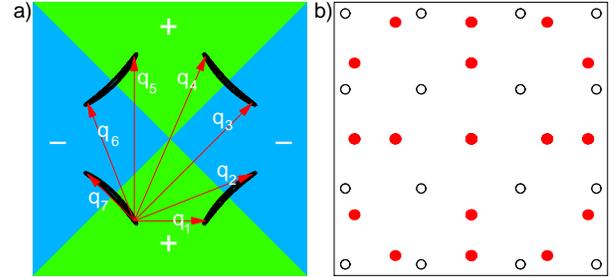}
\caption{(Color online) a) Contours of constant quasiparticle energy $\omega=0.1t$ in a $d$-wave
superconductor in the first Brillouin zone. We use standard tight binding
lattice model \cite{PBF1} with $t'=-0.3t$, $\Delta_0=0.2t$ and $\mu=-t$.
b) Positions of the quasiparticle interference peaks resulting from the octet
model downfolded to the first
Brillouin zone. Peaks enhanced (suppressed) by the magnetic field are marked
by solid (open) circles.}
\label{fig1}
\end{figure}

In this Communication we take a first step in this direction by studying theoretically
the effect
of relatively weak magnetic fields (such that the sample remains in the
superconducting state) on the quasiparticle interference patterns observed
in FT-STS. We find that the field causes strong enhancement
of a {\em subset} of the interference peaks that are observed in zero field.
The pattern of enhancement, illustrated in Fig.\ \ref{fig1}, is closely
related to the $d$-wave symmetry of
the superconducting order parameter and agrees with the recent measurements
performed by Hanaguri {\em et al.} \cite{Hanaguri}. This agreement exemplifies
our level of understanding the the quasiparticle dynamics in cuprates and lays
foundation for the future studies in much stronger fields.

In our subsequent discussion we focus on the quantity $Z(\bk,\omega)$
measured recently by Hanaguri {\em et al.}  \cite{Hanaguri,hanaguri1}
defined as the spatial Fourier transform of the ratio
\begin{equation}\label{z1}
Z(\br,\omega)={g(\br,\omega)\over g(\br,-\omega)},
\end{equation}
where $g(\br,\omega)$ is the tunneling conductance $dI/dV$ measured at
point $\br$ of the sample at bias $\hbar\omega$. A key advantage of considering
the ratio $Z(\br,\omega)$  is that the unknown tunneling matrix element
connecting $g(\br,\omega)$ to the local density of states $n(\br,\omega)$
drops
out (provided that it is a slowly varying function of $\omega$) leaving behind
the ratio of the local density of states which contains information on the
intrinsic electronic state of the system.

There are two very interesting aspects of the above measurements \cite{Hanaguri}:
(i) The quasiparticle interference patterns in $Z(\bk,\omega)$ are even
 clearer and more striking than those observed in $g(\bk,\omega)$ in the
same sample, and
(ii) the patterns in $Z(\bk,\omega)$ are sensitive to the applied uniform
magnetic field in the range 0 to 10T. More specifically, with the increasing
field intensities of various interference peaks vary in a very specific way.
In what follows we formulate a theory of this field-induced variation.

The local density of states (LDOS) in a material can be decomposed into two
parts
\begin{equation}\label{n1}
n(\br,\omega)=n_0(\omega)+ \delta n(\br,\omega),
\end{equation}
The first part is uniform in space, reflecting the physics of a perfectly
homogeneous native material and, for a $d$-wave SC is a V-shaped function
of $\omega$.  The second part describes inhomogeneity due to disorder.
As discussed extensively in the literature, the structure of the quasiparticle
excitation spectra together with the BCS coherence factors cause the Fourier
transform of $\delta n(\br,\omega)$ (which we refer to hereafter as FT-LDOS) to
comprise a collection of sharp peaks \cite{Wang,Capriotti,PBF1}.
The location of these peaks and their dispersion as a function of $\omega$ can
be understood from a heuristic octet model \cite{Hoffman} based on a set of
eight points in the Brillouin zone at the tips of the banana-shaped contours of
constant energy illustrated in Fig.\ \ref{fig1}(a). A peak in FT-LDOS will appear
at momentum $\bq_i$ if it connects any two of the octet points.

To gain theoretical insight into the structure of $Z(\br,\omega)$ we now
substitute Eq.\ (\ref{n1}) into (\ref{z1}) and expand to leading order
in $\delta n$
\begin{equation}\label{z2}
Z(\br,\omega)\simeq Z_0(\omega)\left[1+{\delta n(\br,\omega)\over n_0(\omega)}
- {\delta n(\br,-\omega)\over n_0(-\omega)}\right],
\end{equation}
where $Z_0(\omega)=n_0(\omega)/n_0(-\omega)$. Eq.\ (\ref{z2}) should be an
excellent approximation as long as $|\delta n(\br,\omega)|\ll |n_0(\omega)|$,
a condition well satisfied for the data under consideration \cite{hanaguri1}.
We are interested in the spatially varying part $\delta Z(\br,\omega)$ of the
above expression. It is useful to recast it in the following way:
\begin{equation}\label{z3}
\delta Z(\br,\omega)= C_1(\omega)\delta n_{\rm e}(\br,\omega)
+C_2(\omega)\delta n_{\rm o}(\br,\omega),
\end{equation}
where $\delta n_{\rm e(o)}$ represents the part of $\delta n$ even (odd) in
$\omega$ and $C_{1(2)}$ are $\br$-independent functions of frequency.

The above even/odd decomposition facilitates the following key observation.
As pointed out by Chen {\em et al.} \cite{chen1}, for a strictly {\em particle-hole
symmetric} system, $\delta n_{\rm o}$ originates exclusively from the scattering
in the particle-hole channel (i.e.\ ordinary potential scattering) while
$\delta n_{\rm e}$ comes from scattering in the particle-particle channel (i.e.\
modulation in the SC order parameter). Cuprates are of course not strictly
p-h symmetric but nevertheless they are sufficiently close to the p-h symmetric
situation that the above classification holds to a very good approximation
 \cite{chen1}. In the absence of the magnetic field disorder in the sample gives
rise to ordinary potential scattering as well as off-diagonal scattering
caused by local suppression of the SC gap amplitude by pair-breaking impurities.
Thus, $\delta n$ has in general both even and odd contributions, even in the
strictly p-h symmetric case.

When the magnetic field is applied to the sample in excess of the lower critical
field $H_{c1}$ the sample enters the mixed state and Abrikosov vortices appear.
If there is significant randomness in the Abrikosov lattice then
vortices cause {\em additional quasiparticle scattering} due to (i) the
suppression of the order parameter in the vortex core and (ii) the superflow
associated with the screening currents outside the cores. These effects
both contribute to scattering in the {\em particle-particle channel} and
thus predominantly affect $\delta n_{\rm e}$. Ordinary potential scattering, by
contrast, should be largely unaffected by the magnetic field. We thus
expect magnetic field to {\em enhance}  $\delta n_e$ but leave $\delta n_{\rm o}$
largely unchanged.

In the following we shall explicitly calculate $\delta n(\br,\omega)$ caused
by the order parameter suppression in the vortex core, which we believe is the
dominant effect of the magnetic field. We verify that it is indeed predominantly
even in frequency in the vicinity of the p-h symmetric point and analyze
in detail the spatial structure of the resulting interference pattern reflected
in $Z(\bk,\omega)$.

The local density of states in  a superconductor is given by
$n(\br,\omega) =-{1\over \pi} {\rm Im}
G_{11}(\br,\br;\omega+i\delta)$ where $G(\br,\br';\omega)$ is the full electron
Green's function, a $2\times 2$ matrix in the Nambu-Gorkov space.
For {\em weak} impurity scattering the FT-LDOS can be calculated using the
Born approximation \cite{Capriotti}
\begin{eqnarray}\label{n2}
\delta n(\bq) =-{1\over \pi} V_\alpha(\bq){\rm Im}\sum_{\bk}
[G^0(\bk)\sigma_{\alpha}G^0(\bk-\bq)]_{11},
\end{eqnarray}
where $\sigma_\alpha$ are Pauli matrices in the Nambu space, $V_\alpha(\bq)$ is
the Fourier transform of the random impurity potential in the charge
($\alpha=3$) and spin  ($\alpha=0$) channel. The $\bk$ summation extends over
the first Brillouin zone and the frequency arguments have been suppressed for
brevity. $G^0$ denotes the unperturbed Green's function
\begin{equation}\label{g0}
G^0(\bk,i\omega) = {1 \over \omega^2 + E_k^2 }
\begin{pmatrix} i\omega + \epsilon_k & \Delta_k \\
 \Delta_k & i\omega-\epsilon_k
\end{pmatrix}
\end{equation}
with $\epsilon_{\bk}$ the band energy measured from the Fermi surface,
$\Delta_{\bk}$ the gap function and
$E_{\bk} = \sqrt{\epsilon_{\bk}^2+\Delta_{\bk}^2}$.

 The LDOS modulations
$\delta n(\bk,\omega)$ due to impurity scattering have been extensively
studied \cite{Wang,Capriotti,PBF1,PBF2,nunner1}  based on Eq.\ (\ref{n2}) as
well as more accurate t-matrix calculations. Comparison
to a series of atomic resolution FT-STS
data \cite{Hoffman,mcelroy1,howald1} shows good qualitative agreement in terms of
peak positions and dispersions.

When considering scattering off of spatial modulations of the SC order parameter
such as those occurring near the vortex core, one might expect that a formula
just like Eq.\ (\ref{n2}) but with off-diagonal Pauli matrices ($\alpha=1,2$)
should be applicable. This would indeed be the case for a simple $s$-wave
superconductor. In the case of a $d$-wave order parameter the situation is
slightly more complicated \cite{PBF2}. This is related to the fact that the
$d$-wave order parameter is most naturally described as living on the
{\em bonds} of the underlying square lattice. Correspondingly,
a point-like perturbation will be a gap modulation $\delta\Delta_i$ that affects
four bonds emanating from a single site $\br_i$.  A general gap modulation can
be thought of as a sum of these point-like modulations.

To formulate this we consider a perturbation described by the Hamiltonian
\begin{equation}\label{h1}
\delta{\cal H} = {1\over 2}\sum_{i,\delta}\delta\Delta_i\chi_\delta
[c_\uparrow(\br_i) c_\downarrow(\br_i+\hat\delta)
-c_\downarrow(\br_i) c_\uparrow(\br_i+\hat\delta)+{\rm h.c.}]
\end{equation}
where $c_\sigma(\br)$ represent the electron annihilation operators,
$\hat\delta=\pm\hat x,\pm\hat y$, and $\chi_{\delta}=1$ for $x$-bonds and $-1$
for $y$-bonds. For simplicity we consider $\delta\Delta_i$ real. If we define
the usual Nambu spinor operator $\psi(\br_i)=[c_\uparrow(\br_i),c_\downarrow^\dagger(\br_i)]^T$ we may write
$\delta{\cal H} = \sum_{\bk,\bk'}\psi_{\bk}^\dagger V_{\bk\bk'}\psi_{\bk'}$
with
\begin{equation}\label{vkk}
V_{\bk\bk'}=\sigma_1\delta\Delta_{\bk-\bk'}(\chi_{\bk}+\chi_{\bk'})
\end{equation}
 and
$\chi_{\bk}=\cos{k_x}-\cos{k_y}$ the Fourier transform of $\chi_\delta$. Had
we allowed $\delta\Delta_i$ to have imaginary part there would be an
additional component of $V_{\bk\bk'}$ proportional to $\sigma_2$.

Within the Born approximation we thus have the following vortex-induced LDOS
modulation
\begin{equation}\label{n3}
\delta n(\bq) =-{1\over \pi} {\rm Im}\sum_{\bk}
[G^0(\bk)V_{\bk,\bk-\bq}G^0(\bk-\bq)]_{11}.
\end{equation}
If we now identify $\delta\Delta_{\bq}$ with $V_1(\bq)$ we see that indeed this
result has the form of Eq.\ (\ref{n2}), except for the factor
$(\chi_{\bk}+\chi_{\bk-\bq})$ implied by Eq.\ (\ref{vkk}) that reflects
the $d$-wave symmetry of the order parameter.
Using Eqs.\ (\ref{g0}) and (\ref{vkk}) we may further rewrite
$\delta n(\bq,\omega) = -{1\over \pi}
\delta\Delta_{\bq}{\rm Im}\Lambda(\bq,\omega+i\delta)$ with
\begin{equation}\label{lam1}
\Lambda(\bq,i\omega)=\sum_{\bk}(\chi_+ + \chi_- )
{(i\omega +\epsilon_+ )\Delta_- + (i\omega +\epsilon_- )\Delta_+ \over
(\omega^2+E_{+}^2 )(\omega^2+E_{-}^2 )},
\end{equation}
and $\epsilon_\pm =\epsilon_{\bk\pm\bq/2} $ etc.

In the particle-hole symmetric limit, the terms with $\epsilon_{\pm}$ vanish upon
the momentum summation.  To see this note that in this limit the band energy
has the property $\epsilon_{\bk+\bQ}=-\epsilon_{\bk} $ for $\bQ=(\pi,\pi)$. Since
$\chi_{\bk}$ and $\Delta_{\bk}$ also share this property (irrespective of p-h
symmetry) it follows that shifting the summation variable by $\bQ$ reverses
the sign of $\epsilon_{\pm}$, which therefore vanish in
the sum.  In the p-h symmetric case we are thus left with
\begin{equation}\label{lam2}
\Lambda(\bq,i\omega) ={1\over\Delta_0}\sum_{\bk}
{ i \omega(\Delta_+ + \Delta_- )^2 \over (\omega^2 + E_{+}^2)(\omega^2 + E_{-}^2)},
\end{equation}
where we expressed the gap function as $\Delta_{\bk} =\Delta_0\chi_{\bk} $.
As will be evident shortly, the remaining expression is even in $\omega$
when analytically  continued to real frequencies.

We now wish to examine the effect of this contribution on the quasiparticle
interference peaks that are predicted by the octet model. To this end it is
useful to perform analytical continuation and explicitly evaluate
$-{1\over \pi} {\rm Im}\Lambda(\bq,\omega+i\delta)$ at wavectors
$\bq_{ij}=\bQ_i - \bQ_j $, where $\bQ_i$ are the octet vectors. This yields
\begin{equation}\label{lam3}
{1\over\Delta_0}\sum_{\bk} \delta(|\omega|-E_{\bk}){\cal P}
{(\Delta_{\bk-\bq} + \Delta_{\bk} )^2 \over (E_{\bk-\bq}^2- E_{\bk}^2)},
\end{equation}
where ${\cal P}$ denotes the principal part. It is clear that for $\bq=\bq_{ij}$
the largest contribution to the sum comes from the vicinity of $\bk=\bQ_i,\bQ_j$.
Near these points the denominator approaches zero but numerator is a slowly
varying function of $\bk$. We may thus approximate the latter by its value at
$\bk=\bQ_i,\bQ_j$ and take it outside of the sum. We thus obtain
\begin{equation}\label{lam4}
-{1\over \pi} {\rm Im} \Lambda(\bq_{ij}) \approx (\Delta_i+\Delta_j)^2
{1\over\Delta_0} \sum_{\bk} {\cal P}
{ \delta(|\omega|-E_{\bk})\over (E_{\bk-\bq_{ij}}^2- E_{\bk}^2)},
\end{equation}
where $\Delta_i$ denotes $\Delta_{\bq}$ evaluated at the octet vector
$\bQ_i$.

The above Eq.\ (\ref{lam4}) has some remarkable implications and represents
our main result. Most importantly it implies that in the p-h symmetric case
the effect of the applied magnetic field on the octet vectors can be summarized
as
\begin{equation}\label{n4}
\delta n_{\rm e}(\bq_{ij},\omega) \sim (\Delta_i+\Delta_j)^2 {\cal K}_{ij}(\omega).
\end{equation}
Here ${\cal K}_{ij}(\omega)$ denotes the sum in Eq.\ (\ref{lam4}) and can be
shown to represent a positive quantity whose precise value for a given
frequency and vector $\bq_{ij}$ depends on the details of the
underlying band structure and is thus non-universal. The factor
$(\Delta_i+\Delta_j)^2$ is, by contrast, universal and depends only on the
symmetry
of the SC order parameter. Since the octet points lie on the Fermi surface
of the underlying normal metal it is easy to see that
$|\Delta_i|=|\omega|$ for all $i$'s. The sign of $\Delta_i$, however depends
on the position of $\bQ_i$ in the Brillouin zone as illustrated in Fig.\
\ref{fig1}(a). It then follows that
\begin{equation}\label{n5}
\delta n_{\rm e}(\bq_{ij},\omega) \sim \biggl\{
\begin{array}{cc}
4\omega^2 {\cal K}_{ij}(\omega), & {\rm if}\ \sgn{\Delta_i}=\sgn{\Delta_j}\\
0,   & {\rm if}\ \sgn{\Delta_i}\neq\sgn{\Delta_j}\\
\end{array}
\end{equation}
Thus, remarkably, we find that only those interference peaks will be enhanced
by applied magnetic field
whose wavectors $\bq_{ij}$ connect octet points in the regions of the
Brillouin zone
with the same sign of the gap function $\Delta_{\bk}$, denoted by $+/+$ and
$-/-$ in Fig.\ \ref{fig1}(a). These are $\bq_1$, $\bq_4$, and $\bq_5$.
The remaining  $+/-$ peaks will be to leading order unaffected. The resulting
pattern
is illustrated in Fig.\  \ref{fig1}(b). We remark that these are precisely the
peaks observed to be enhanced in the
experiments by Hanaguri {\em et al.} \cite{Hanaguri}.

If the system breaks the p-h symmetry, as is generally the case in cuprates at
finite doping concentration, then the above conclusion cannot be formulated
as a precise symmetry statement. However, as long as the  p-h symmetry
breaking remains relatively weak, as is the the case in cuprates close to half
filling, our result Eq.\ (\ref{n5}) will hold to a good approximation:
the $+/+$ and $-/-$ peaks will be significantly enhanced while $+/-$ will
be affected only slightly. This can be seen by analyzing Eq.\ (\ref{lam1}).
Even when  p-h symmetry is weakly broken we can still write it as
(\ref{lam2}) plus a correction that will be {\em odd} in frequency. This
correction will be (i) small compared to the leading term by factors of
$\mu/t$ and $t'/t$, where $\mu$ is the chemical
potential and $t$, $t'$ are nearest and next-nearest neighbor hopping amplitudes,
and (ii) will contain a factor $(\Delta_+ +\Delta_-)$ which will make it
very small for the $+/-$ peaks, as before.

\begin{figure}[t]
\includegraphics[width = 8cm]{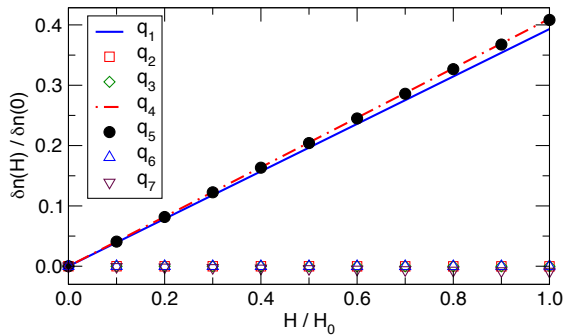}
\caption{(Color online) Enhancement of FT-LDOS peaks by magnetic field
at octet vectors $\bq_i$   modeled by Eq.\ (\ref{lam1}) with the band structure
as in Fig.\ \ref{fig1}. In the calculation the field $H$ is
represented by the vortex core scattering rate $V_1$ and $H_0$ is chosen such
that $H/H_0$=$V_1/V_3$. The data are normalized to the $H=0$ value which we model
by $V_1=0$ and $V_3>0$, i.e. charged impurities only.
}
\label{fig2}
\end{figure}

In order to ascertain the validity of our above conclusions
 we have evaluated the sums indicated in Eq.\ (\ref{lam1})
numerically for band structures with realistic parameters. This
is illustrated in Fig.\ \ref{fig2}.
We have also analyzed
these expressions within the nodal approximation along the lines
of Refs.\ \onlinecite{PBF1, PBF2}. These considerations confirm that
Eq. (\ref{n5}) is an excellent proxy for Eq.\ (\ref{lam1}) when the p-h
symmetry is present as well as when it is weakly violated, as in the
superconducting state of cuprates. Specifically, we find that magnetic field
enhances the peaks at $\bq_1$, $\bq_4$, and $\bq_5$, while the remaining peaks
are essentially unaffected. This pattern of enhancement remains surprisingly
robust even when the p-h symmetry is significantly violated as in
Fig.\ \ref{fig2}.

Hanaguri {\em et al.} \cite{Hanaguri} report that the $+/+$ and $-/-$ are
enhanced by the applied magnetic field and the $+/-$ peaks are in fact
reduced in amplitude. This can be reconciled with our theoretical prediction
when we recall that our model explicitly treats only one aspect of the field,
namely the suppression of the order parameter in the vortex cores. Magnetic
field also generates superflow which is known to produce a Doppler shift
 \cite{volovik1} and a more subtle Berry phase \cite{ft1} effect on the
quasiparticle wavefunctions. Since these are both long-range, non-local effects,
their impact on the quasiparticle interference patterns is significantly more
difficult to compute. It appears to us likely, however, that the additional
phases acquired by the quasiparticles as they propagate on the background of the
random vortex array will tend to scramble the interference patterns and therefore
{\em suppress} the peaks. We thus hypothesize that a combination of this
suppression of all peaks and the enhancement of $+/+$ and $-/-$ peaks
due to the vortex core scattering will lead to the pattern observed
in experiment \cite{Hanaguri}.

Our results here underscore once again the importance of the quasiparticle
coherence factors \cite{PBF1} for the tunneling interference spectroscopy.
Indeed, the pattern of the peak enhancement by magnetic field found here is determined
solely by the coherence factors. Their presence, manifested in the peak-like FT-STS patterns, indicates pairing even in magnetic field.
Our results, in conjunction with experimental
data \cite{Hanaguri}, also illustrate the remarkable sensitivity of the FT-STS
technique to relatively modest magnetic fields up to 10T. This suggests
good prospects of FT-STS for unraveling the mystery of the normal state
reached at higher fields or temperatures.

The authors are indebted to J.C. Davis and T. Hanaguri for discussing their data
prior to publication and to P. Coleman, E.-A. Kim and L. Taillefer for
illuminating discussions. This work
was supported by NSERC, CIFAR and the Killam Foundation.



\begin{thebibliography}{999}
\bibitem{franz0} M. Franz, Science {\bf 305}, 1410 (2004).
\bibitem{taillefer1} N. Doiron-Leyraud, Cyril Proust, David LeBoeuf, Julien Levallois, Jean-Baptiste Bonnemaison, Ruixing Liang, D. A. Bonn, W. N. Hardy, Louis Taillefer, Nature {\bf 447}, 565 (2007).
\bibitem{kanigel1} A. Kanigel, M. R. Norman, M. Randeria, U. Chatterjee, S. Souma, A. Kaminski, H. M. Fretwell, S. Rosenkranz, M. Shi, T. Sato, T. Takahashi, Z. Z. Li, H. Raffy, K. Kadowaki, D. Hinks, L. Ozyuzer, J. C. Campuzano, Nature Phys. {\bf 2} 447 (2006).
\bibitem{shen1} W. S. Lee, I. M. Vishik, K. Tanaka, D. H. Lu, T. Sasagawa, N. Nagaosa, T. P. Devereaux, Z. Hussain, Z.-X. Shen, Nature {\bf 450}, 81 (2007).
\bibitem{fischer1} \O. Fischer, M. Kugler, I. Maggio-Aprile, and C. Berthod,
Rev. Mod. Phys. {\bf 79}, 353 (2007).
\bibitem{Hoffman}J. Hoffman, K. McElroy, D.-H. Lee, K.M. Lang, H. Eisaki, S. Uchida, and J.C. Davis,
Science {\bf 297}, 1148 (2002).
\bibitem{mcelroy1}K. McElroy,
R.W. Simmonds, J.E. Hoffman, D.-H. Lee, J. Orenstein, H. Eisaki, S. Uchida, and J.C. Davis,
Nature {\bf 422}, 592 (2003).
\bibitem{howald1} C. Howald, H. Eisaki, N. Kaneko, M. Greven, and A. Kapitulnik,
\prb {\bf 67}, 014533 (2003).
\bibitem{Hanaguri} T. Hanaguri {\it et al.}, unpublished.
\bibitem{hanaguri1} T. Hanaguri, Y. Kohsaka, J. C. Davis, C. Lupien, I. Yamada, M. Azuma, M. Takano, K. Ohishi, M. Ono, H. Takagi, Nature Phys. {\bf 3}, 865 (2007).
\bibitem{Wang} Q.-H. Wang and D.-H. Lee \prb {\bf 67}, 20511(R) (2003)
\bibitem{Capriotti}L. Capriotti, D.J. Scalapino, and R.D. Sedgewick, Phys. Rev. B {\bf 68}, 014508 (2003).
\bibitem{PBF1} T. Pereg-Barnea and M. Franz, \prb {\bf 68}, 180506(R) (2003).
\bibitem{chen1} H.-D. Chen, O. Vafek, A. Yazdani, and S.-C. Zhang,
\prl {\bf 93}, 187002 (2004).
\bibitem{PBF2}T. Pereg-Barnea and M. Franz, Int.\ J.\ Mod.\ Phys.\ B {\bf 19},
731 (2005).
\bibitem{nunner1}  Tamara S. Nunner, Wei Chen, Brian M. Andersen, Ashot Melikyan, and P. J. Hirschfeld, \prb {\bf 73}, 104511 (2006).
\bibitem{volovik1} G.E. Volovik, JETP Lett. {\bf 58}, 469 (1993).
\bibitem{ft1} M. Franz and Z. Te\v{s}anovi\'{c}, \prl {\bf 84}, 554 (2000).

\end{thebibliography}
\end{document}